\documentclass[11pt,aps,amsmath,amssymb,nofootinbib,notitlepage,longbibliography]{revtex4-1}
\usepackage{amsmath,amssymb}
\usepackage{slashed}
\usepackage{graphicx}
\usepackage{bm}
\usepackage[top=1.0in,bottom=1.0in,left=1.0in,right=1.0in]{geometry}
\usepackage[colorlinks,linkcolor=blue,citecolor=blue]{hyperref}

\newcommand{\be}{\begin{equation}}
\newcommand{\ee}{\end{equation}}
\newcommand{\bea}{\begin{eqnarray}}
\newcommand{\eea}{\end{eqnarray}}
\newcommand{\ba}{\begin{eqnarray}}
\newcommand{\ea}{\end{eqnarray}}

\begin{document}

\title{Nucleon mass radii and distribution
:\\ Holographic QCD, Lattice QCD and GlueX data}

\author{Kiminad A. Mamo}
\email{kmamo@anl.gov}
\affiliation{Physics Division, Argonne National Laboratory, Argonne, Illinois 60439, USA}

\author{Ismail Zahed}
\email{ismail.zahed@stonybrook.edu}
\affiliation{Center for Nuclear Theory, Department of Physics and Astronomy, Stony Brook University, Stony Brook, New York 11794--3800, USA}

\begin{abstract}
We briefly review and expand our recent analysis for all three invariant A,B,D gravitational form factors of the nucleon in holographic QCD. 
They compare well to the gluonic gravitational form factors recently measured using lattice  QCD  simulations. The  holographic
A-term is fixed by the tensor $T=2^{++}$ (graviton) Regge trajectory, and the D-term by the difference between the tensor $T=2^{++}$ (graviton)
and scalar $S=0^{++}$ (dilaton)  Regge trajectories.
 The B-term is null in the absence of a tensor coupling to a Dirac fermion in bulk. 
 A first measurement of the tensor form factor A-term  is already accessible  using the current
 GlueX data, and therefore the tensor gluonic mass radius, pressure and shear inside the proton, thanks
 to holography. The holographic A-term and D-term can be expressed exactly in terms of  harmonic numbers.
The   tensor mass radius  from the holographic threshold is found to be $\langle r^2_{GT}\rangle \approx (0.57-0.60\,{\rm fm})^2$, 
 in agreement with  $\langle r^2_{GT}\rangle \approx (0.62\,{\rm fm})^2$ as extracted from the  overall
 numerical lattice   data, and empirical GlueX data. The scalar mass radius is found to be slightly larger
 $\langle r^2_{GS}\rangle \approx (0.7\,{\rm fm})^2$.
 \end{abstract}

\maketitle

%%%%%%%%%%%%%%%%%%%%%%%%%%%%%%%%%%%%%%%%
\section{Introduction}

A persistent and fundamental question in   physics is about the   origin of mass in the nucleon, and therefore in all
visible hadronic mass in the Universe.  Where does it come from, and how is it distributed inside the nucleon?
Unlike the Higgs in electroweak theory which is at the origin of the leptonic and current 
quark masses, most of the hadronic mass of the visible Universe stems  from QCD, a theory with almost no mass
~\cite{Wilczek:2012sb,Roberts:2021xnz} (and references therein).

The dual quantum breaking of conformal symmetry and the spontaneous breaking of chiral symmetry in QCD 
are at the origin of this mass without mass. These two fundamental phenomena are  tied by strong
topological fluctuations in the QCD vacuum: eternal tunneling events between gauge vacuua with different 
winding numbers, also known as  instantons and anti-instantons~\cite{Shuryak:2018fjr,Zahed:2021fxk} (and references therein).

While decisive understanding of these two phenomena has been achieved theoretically and numerically
using lattice QCD simulations~\cite{Leinweber:1999cw,Biddle_2020},  empirical measurements to support this understanding is only now emerging at 
current and dedicated electron machines~\cite{Hafidi:2017bsg,Ali:2019lzf,Meziani:2020oks,Anderle:2021wcy}. 
Recently, the GlueX collaboration~\cite{Ali:2019lzf} at JLAB has reported threshold data for
photo-production of charmonium $J/\Psi$ that may start to lift the lid on some of these fundamental questions.
Indeed, near threshold the elastic production of a heavy vector meson is likely to proceed mostly through gluons or more precisely 
tensor glueballs, as is the case way above threshold through the strong Pomeron exchange  in the diffractive regime~\cite{Costa:2013uia,Lee:2018zud}.

In a recent analysis of the GlueX data using a holographic construction, we have shown~\cite{Mamo:2019mka} 
that the threshold differential cross section is only sensitive to the tensor gravitational form factor, and suggested
that  this tensor  form factor or A-term  is extractable from the current data under a minimal but universal set of holographic assumptions. This allows for a first extraction
of the tensor mass radius among other things. Remarkably, the holographic construction ties the A- and D-gravitational
form factors, thereby allowing for the extraction of the gluonic pressure and shear inside the proton. For completeness,
we note the  holographic discussion regarding the extraction of the gluon condensate  in the proton using the GlueX data in~\cite{Hatta:2018ina}.

In section~\ref{ABC} we briefly review and expand our  arguments for the holographic A,B,D invariant gravitational   
form factors for the proton, extract the tensor mass radius, and compare the results to the most recent lattice data. In section~\ref{PRESSURE} we use
the holographic relationship between the A- and D-term to analyze the gluonic pressure and shear inside the proton.
The pressure inside the proton results from a delicate balance between the repulsive tensor  glueball  contribution at short distances,
and the attractive scalar glueball contribution at large distances.  An estimate of the scalar mass radius is made.
In section~\ref{GlueXDATA} we show how to use the current GlueX data to extract empirically the A-form factor.
The result   is  in remarkable agreement with  the holographic result and lattice data.
Our conclusions are in section{\ref{CONCLUSION}.

\section{Gravitational form factors}~\label{ABC}

The standard decomposition of the  energy-momentum form factor in a nucleon state is~\cite{Pagels:1966zza,Carruthers:1971uy,Polyakov:2018zvc}

\be
\label{A1}
\left<p_2|T^{\mu\nu}(0)|p_1\right>=\overline{u}(p_2)\left(
A(k)\gamma^{(\mu}p^{\nu)}+B(k)\frac{ip^{(\mu}\sigma^{\nu)\alpha}k_\alpha}{2m_N}+C(k)\frac{k^\mu k^\nu-\eta^{\mu\nu}k^2}{m_N}\right)u(p_1)\,,
\ee
with  $a^{(\mu}b^{\nu)}=  \frac 12 (a^\mu b^\nu+a^\nu b^\mu)$, 
 $k^2=(p_2-p_1)^2=t$, $p=(p_1+p_2)/2$  and the normalization $\overline u u=2m_N$. 
(\ref{A2}) is conserved and tracefull. Throughout, $D(k)=4C(k)$ will be used interchangeably.
In holography,  (\ref{A1}) sources the metric fluctuations in bulk, 
\be
g_{MN}(z)\rightarrow g_{MN}(z)+h_{MN}(x,z)
\ee
with line element $ds^2=g_{MN}(z)dx^Mdx^N$ in a 5-dimensional anti-deSitter space or AdS$_5$,  in the double limit of large $N_c$ and strong gauge coupling~\cite{Nastase:2007kj}
(and references therein).
The form factors in (\ref{A1}) follow from the coupling  of the irreducible representations of the metric fluctuations  $h_{\mu\nu}$,  to a bulk Dirac fermion 
with chiral components $\psi_{L,R}$.
 The bulk metric fluctuations can be decomposed  in terms of the $2\oplus 1\oplus 0$ invariant tensors~\cite{Kanitscheider:2008kd}

\be
h_{\mu\nu}(k,z)=\bigg[\epsilon_{\mu\nu}^{TT}h(k,z)+k_{\mu} k_{\nu}H(k,z)\bigg]+\bigg[k_{\mu}A^{\perp}_{\nu}(k,z)+k_{\nu}A^{\perp}_{\mu}(k,z)\bigg]+\bigg[\frac{1}{3}\eta_{\mu\nu}f(k,z)\bigg]
\ee
which is the spin-2 made of the transverse-traceless part $h$  plus the  longitudinal-tracefull part $H$, the spin-1 made of the transverse vector $A_\mu^\perp$,
and the spin-0 tracefull part $f$.
%Using a Witten-diagram and the holographic dictionary, the result is~\cite{Abidin:2009hr,Mamo:2019mka}

\subsection{A-term}

To determine the A-term, we 
contract   the  energy-momentum form factor (\ref{A1}) with a spin-2 transverse-traceless polarization tensor $\epsilon_{\mu\nu}^{TT}$,

\bea
\label{EMT2}
\left<p_2|\epsilon_{\mu\nu}^{TT}T^{\mu\nu}(0)|p_1\right> &=&\overline{u}(p_2)\left(
A(k)\epsilon_{\mu\nu}^{TT}\gamma^{\mu}p^{\nu}\right)u(p_1)\nonumber\\
&=&\overline{u}(p_2)\frac{\delta}{\delta h_0}\left(\frac{\tilde{A}}{2}\int dz\sqrt{g}\,e^{-\phi(z)}z\,\big(\psi_R^2(z)+\psi_L^2(z)\big)h(k,z)\times \epsilon_{\mu\nu}^{TT}\gamma^\mu p^\nu \right)u(p_1)\nonumber\\
&=&\overline{u}(p_2)\left(\frac{\tilde{A}}{2}\int dz\sqrt{g}\,e^{-\phi(z)}z\,\big(\psi_R^2(z)+\psi_L^2(z)\big)\chi(k,z)\times \epsilon_{\mu\nu}^{TT}\gamma^\mu p^\nu \right)u(p_1)\,,\nonumber\\
\eea
The last two lines follow from a tree level Witten-diagram and the holographic dictionary in the soft wall construction as detailed in~\cite{Nastase:2007kj,Abidin:2009hr,Mamo:2019mka}.  They correspond to the
 coupling  of the transverse-traceless part of the graviton  $h(k,z)=h_0\chi(k,z)$ with $\chi(k,0)=1$ (dual to $2^{++}$ tensor glueballs) to a Dirac fermion in bulk. 
 More specifically~\cite{Mamo:2019mka}~\footnote{In~\cite{Mamo:2019mka} there is a typo in the argument of the di-gamma function $\psi(x)\rightarrow \psi(1+x)$.}

\bea
\label{A2}
A(k)&=&\frac{\tilde{A}}{2}\int dz\sqrt{g}\,e^{-\phi(z)}z\,\big(\psi_R^2(z)+\psi_L^2(z)\big)\chi(k,z)\nonumber\\
&=&\tilde{A}\left((1-2a_k)(1+a_k^2)+a_k(1+a_k)(1+2a_k^2)\left(H\left(\frac{1+a_k}{2} \right)-H\left(\frac{a_k}{2}\right)\right)\right)\nonumber\\
\eea
with $a_k={-k^2}/8\kappa_N^2$. Here  $H(x)=\psi(1+x)+\gamma$ is the harmonic number or digamma function
 plus Euler number.  The scale $\kappa_N$  follows from the dilaton profile
$\phi(z)=\kappa_N^2 z^2$.
% in the Einstein-Hilbert action in the string frame.
 It is dual to the 
string tension  in QCD.  $A(0)$ is not fixed in holography (1-point function). 
We have defined  $\tilde{A}=A(0)$ to encode $\mathcal{O}(1/N_c)$ corrections coming from 1-loop and higher Witten diagrams for the  transverse-traceless 
tensor part of the energy momentum tensor.   We have checked that (\ref{A2}) is in numerical agreement with a result in~\cite{Abidin:2009hr}  modulo the overall normalization.

\subsection{D-term}

To determine the C-term or  D-term ($D=4C$), we 
contract   the  energy-momentum form factor (\ref{A1}) with  $\frac{1}{3}\eta_{\mu\nu}$, 

\bea
\label{EMT22}
\frac{1}{3}\left<p_2|\eta_{\mu\nu}T^{\mu\nu}(0)|p_1\right> &=&\overline{u}(p_2)\left(
A(k)\frac{m_N}{3}+\frac {k^2}{12m_N}B(k)-\frac{k^2}{m_N}C(k)\right)u(p_1)\nonumber\\
&=&\overline{u}(p_2)\frac{\delta}{\delta f_0}\left(\frac{\tilde{C}}{2}\int dz\sqrt{g}\,e^{-\phi(z)}z\,\big(\psi_R^2(z)+\psi_L^2(z)\big)f(k,z)\times \frac{1}{3}\eta_{\mu\nu}\gamma^\mu  p^\nu \right)u(p_1)\nonumber\\
%&=&\overline{u}(p_2)\left(\frac{\tilde{C}}{2}\int dz\sqrt{g}\,e^{-\phi}z\,\big(\psi_R^2(z)+\psi_L^2(z)\big)\times \frac{1}{3}\eta_{\mu\nu}\frac{p^{\mu}}{m_N} p^\nu \right)u(p_1)\,,
&=&\overline{u}(p_2)\left(\frac{\tilde{C}}{2}\int dz\sqrt{g}\,e^{-\phi(z)}z\,\big(\psi_R^2(z)+\psi_L^2(z)\big)\chi(k, z)\times \frac{m_N}3\right)u(p_1)\nonumber\\
&= &\overline{u}(p_2)\left( A_S (k)\frac{m_N}{3} \right)  u(p_1)
\eea
where the constant $A_S(0)=\tilde{C}$ encodes the $\mathcal{O}(1/N_c)$ corrections coming from 1-loop and higher Witten diagrams for 
the tracefull part of the energy momentum tensor.
In the  line before last in (\ref{EMT22}) we used the coupling  of the tracefull part of the graviton or dilaton  $f(k,z)=f_0-4\phi_0+4\phi_0\chi(k,z)$
with $f_0=4\phi_0$~\cite{Kanitscheider:2008kd} (dual to $S=0^{++}$ scalar glueballs) to a Dirac fermion in bulk,
and again a  tree level Witten-diagram and the holographic dictionary as detailed in~\cite{Nastase:2007kj,Abidin:2009hr,Mamo:2019mka}. 
More specifically~\footnote{In~\cite{Mamo:2019mka} $C(k)$ was taken to be sourced
by the scalar  bulk field $f(k,z)$. Due to mixing of glueball fields in bulk this is an approximation. In fact $f(k,z)$  sources exactly  the scalar dilaton field $T^\mu_\mu$  as we now have shown.}

% to extract the form factor $C(k)$, in the soft wall model as
\bea
\label{A222}
C(k)=
%&&-\frac{1}{3}\frac{m_N^2}{k^2}\left(\frac{\tilde{C}}{2}\int dz\sqrt{g}\,e^{-\phi(z)}z\,\big(\psi_R^2(z)+\psi_L^2(z)\big)-A(k)\right)+\frac 1{12}B(k)\nonumber\\
\frac{1}{3}\frac{m_N^2}{k^2}\bigg(A(k)-A_S(k)\bigg)+\frac 1{12}B(k)\rightarrow\frac{1}{3}\frac{m_N^2}{k^2}\bigg(A(k)-A_S(k)\bigg)
\eea
with $A_S(k)/A(k)=\tilde C/\tilde A$, as the $T=2^{++}$ and $S=0^{++}$ glueballs are degenerate in the present soft wall construction
at large $N_c$ (they have the same anomalous dimension $\Delta_{T,S}=4$).
Since the Pauli-like  form factor $B(k)=0$, as  the coupling of the graviton to the bulk Dirac fermion through 
the spin-connection vanishes,  the rightmost result follows.

The absence of massless modes in the scalar channel requires that $\tilde C=\tilde A$.   This can also be checked by
inserting  (\ref{A222}) in  (\ref{A1}) and tracing,  so that $\left<p|T^{\mu}_{\mu}(0)|p\right>=2\tilde{C}m_N^2$
in any frame. Similarly, the $00$-component of the energy momentum tensor gives $\left<p|T^{00}(0)|p\right> =2 \tilde A p_0^2\rightarrow 2\tilde{A}m_N^2$,
with the rightmost result following in the rest frame only. Both results are expected from Poincare symmetry
$\left<p|T^{\mu\nu}(0)|p\right>=2p^\mu  p^\nu$. Therefore $\tilde A=\tilde C$ can be identified (they are 1 in full QCD,  but  less than 1 in the holographic limit of QCD
which is gluonic in leading order in $1/N_c$)  and (\ref{A222}) vanishes.

At  finite $N_c$, the anomalous dimensions are not equal  with $\Delta_T\neq \Delta_S\neq 4$, e.g.~\cite{Gubser:2008yx}

\be
\Delta_S\rightarrow 4+\bigg(\beta^\prime(\alpha_s)-\frac 2{\alpha_s}\beta(\alpha_s)\bigg)
\ee
corrected by the beta-function $\beta(\alpha_s)$, and 
$C(k)$ does not vanish in general.
Only   $A(0)=A_S(0)=1$ is
required by the  absence of a massless pole and Poincare symmetry. Note that since $m_S<m_T$   the form factor $C(k)$ is in general negative. Also note that
the $ S=0^{++}$ glueball is expected to mix strongly with the scalar sigma-meson at large but finite $1/N_c$.

\subsection{Gluonic gravitational radii}

The tensor gravitational radius following from  (\ref{A1}) is ($k^2=t=-K^2$)

\be
\label{AR}
\langle r^2_{GT}\rangle =-6\bigg(\frac{d{\rm Ln}A(K)}{dK^2}\bigg)_0=\bigg(2-H\bigg(\frac 12\bigg)\bigg)\frac 3{4\kappa_N^2}=\frac {1.04}{\kappa_N^2}
\ee
The $T=2^{++}$ tensor and $0^{++}$ scalar glueballs map to a graviton  and dilaton dual to  $T^{\mu\nu}$ and $T_\mu^\mu$ respectively  in bulk, 
with anomalous dimensions $\Delta_{T,S}=4$ and 5-dimensional squared masses $m_{5,T,S}^2=\Delta_{T,S}(\Delta_{T,S}-4)=0$, 
with Regge trajectories  given by~\cite{Forkel:2007ru,Colangelo:2007if,BoschiFilho:2012xr}

\be
\label{A3}
m_{T,S}^2(n)=8\kappa_N^2\bigg(n+2\bigg)
%=8\kappa_N^2\bigg(n+\frac {\Delta_{T,S}} 2\bigg)=8\kappa_N^2\bigg(n+2\bigg)
\ee
The slope is related to half the  slope of the nucleon (twist $\tau_N=3$) and rho meson radial Regge trajectories

\be
\label{A4}
m_{N}^2(n)=4\kappa_N^2(n+\tau_N -1)\qquad m_{\rho }^2(n)=4\kappa_N^2(n+1)
\ee
Recall that   the  dilaton profile is  $\phi(z)=\kappa_N z^2$ in the DBI action (nucleon and rho) and
$2\phi(z)$ in the Einstein-Hilbert action in the string frame (graviton).

A simultaneous fit to the rho-meson and nucleon radial trajectories (\ref{A4})  is achieved by choosing $\kappa_N\approx 350\,{\rm MeV}$
as in the photoproduction analysis in~\cite{Mamo:2019mka},
which gives $m_N\approx 990\,{\rm MeV}$ and $m_\rho\approx 700\,{\rm MeV}$  and therefore  the degenerate glueball
masses $m_{T}(0)=m_{S}(0)=\sqrt{2}m_N\approx 1386\,{\rm MeV}$. The latters are to be compared to the lattice glueball masses
$m_{2++}\approx  2150\,{\rm MeV}$ and $m_{0++}\approx 1475\,{\rm MeV}$~\cite{Meyer:2004gx}
(see also~\cite{Lucini:2001ej} for slightly heavier glueballs).  The corresponding tensor gravitational radius (\ref{AR}) is

\be
\label{A6}
\langle r^2_{GT}\rangle  \approx (0.57\,{\rm fm})^2
\ee
Conversely, if we fix the  nucleon mass $m_N=940\,{\rm MeV}$ then $\kappa_N\approx 330\,{\rm MeV}$, 
the glueball masses are $m_{T}(0)=m_{S}(0)\approx 1330$ MeV, and the tensor gravitational radius (\ref{AR}) is slightly larger

\be
\label{A6X}
\langle r^2_{GT}\rangle  \approx (0.60\,{\rm fm})^2
\ee
Note  that the  scalar dilaton field $T_\mu^\mu$  in (\ref{EMT22}) is characterized by the scalar form factor $A_S(k)$ ($S=0^{++}$ glueball)  
with an equal mass radius in the strict holographic limit, i.e. $\left<r^2_{GS}\right>=\left<r^2_{GT}\right>$. The scalar radius is slightly larger  at finite $1/N_c$ 
(see (\ref{AZ3}) below).

\begin{figure}[!htb]
\minipage{0.8\textwidth}
  \includegraphics[width=\linewidth]{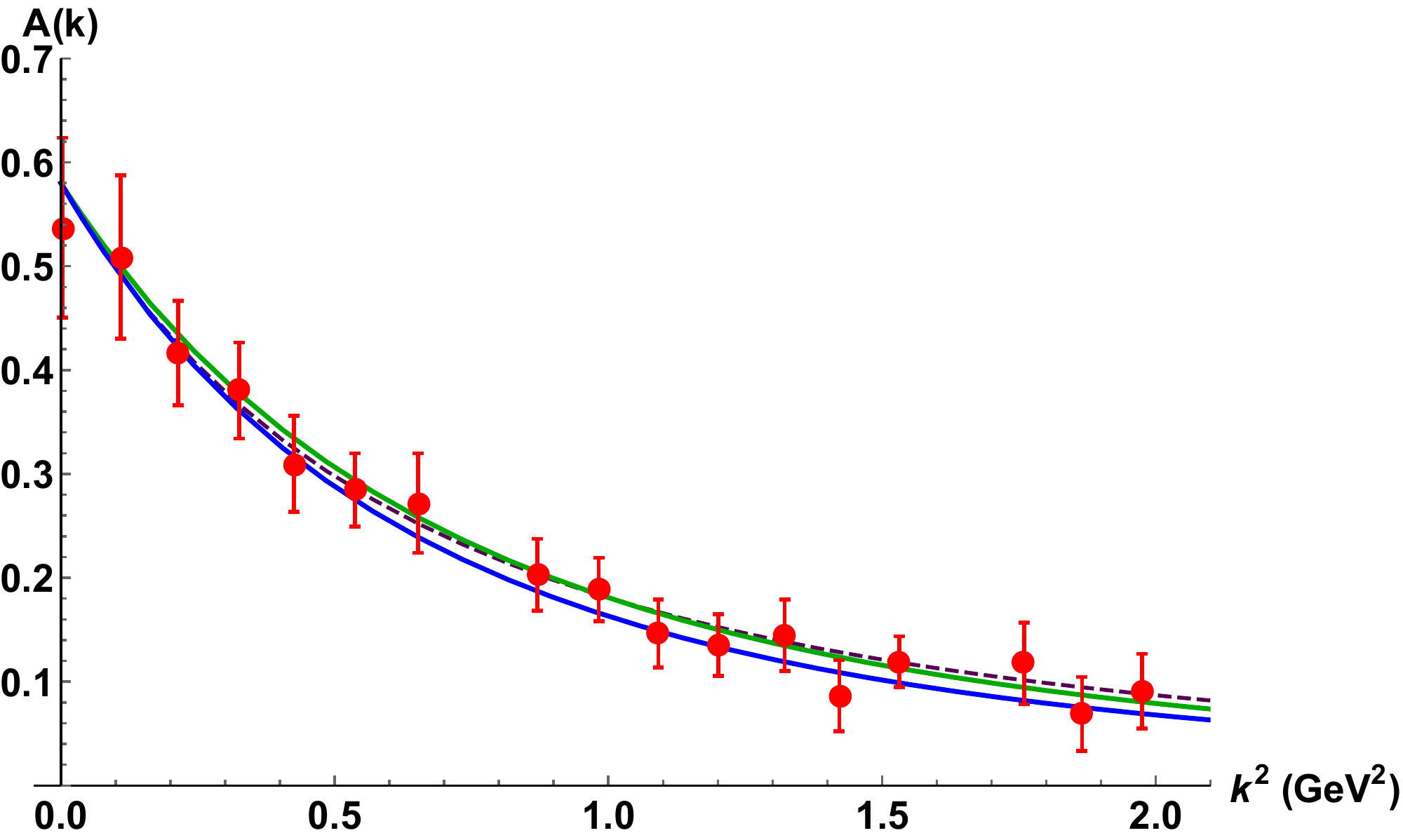}
 % \caption{A really Awesome Image}\label{fig:awesome_image1}
\endminipage\hfill
\minipage{0.8\textwidth}
 \includegraphics[width=\linewidth]{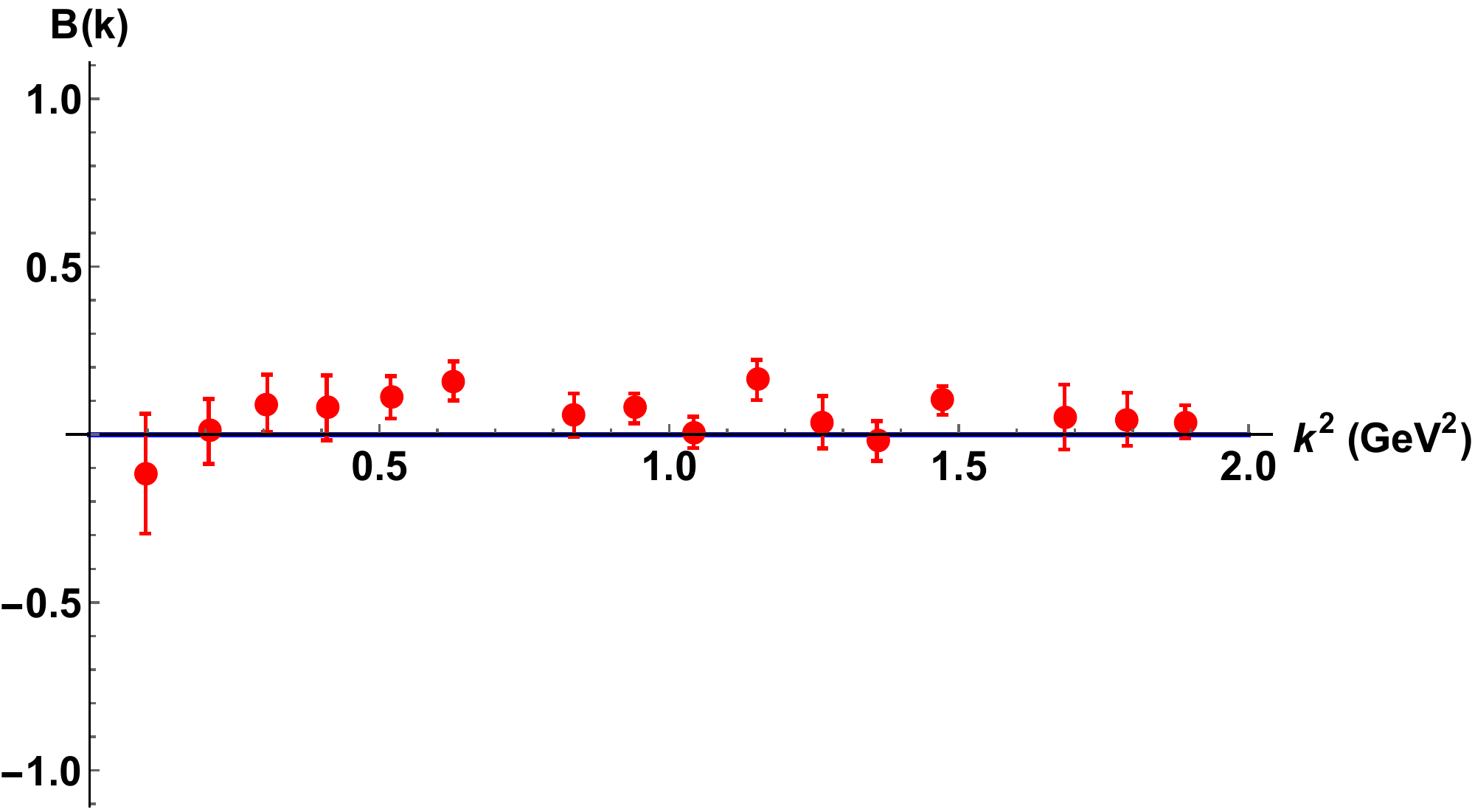}
 % \caption{A really Awesome Image}\label{fig:awesome_image2}
\endminipage\hfill
%\minipage{0.6\textwidth}%
%  \includegraphics[width=\linewidth]{DK.pdf}
 % \caption{A really Awesome Image}\label{fig:awesome_image3}
%\endminipage
 \caption{Holographic  nucleon gravitational form factors:  A(k) 
 (blue-solid  and green-solid curves~(\ref{A2})  
 and purple-dashed curve~(\ref{A7})) and B(k) (blue-solid curve)~\cite{Mamo:2019mka},  versus the unquenched  gluon lattice results  (red-dots)~\cite{Shanahan:2018pib}. See text.}
 % : A(k) top, B(k) middle and C(k) bottom.}
 \label{fig_FF}
\end{figure}

\subsection{Comparison to lattice results}

In the lattice QCD calculation of the gravitational form factors ~\cite{Shanahan:2018pib}, they use the  decomposition of the  energy-momentum form factor 
in a nucleon state  which is traceless and non-conserved

\be
\label{A1lattice}
\left<p_2|T_L^{\mu\nu}(0)|p_1\right>=\overline{u}(p_2)\left(
A(k)\gamma^{[\mu}p^{\nu]}+B(k)\frac{ip^{[\mu}\sigma^{\nu]\alpha}k_\alpha}{2m_N}+C(k)\frac{k^{[\mu} k^{\nu]}}{m_N}\right)u(p_1)\,,
\ee
with  $a_{[\mu}b_{\nu]}=\frac{1}{2}(a_{\mu}b_{\nu}+a_{\nu}b_{\mu})-\frac{1}{4}\eta_{\mu\nu}a_{\alpha}b^{\alpha}$.  (\ref{A1lattice}) cannot
be probed  directly by a metric fluctuation in bulk, since the latter couples only to the conserved energy-momentum tensor.  However, the invariant form 
factors are those of the conserved energy-momentum tensor, so they follow from the latter.

%The nucleon gravitational form factors have been recently analyzed on the lattice~\cite{Shanahan:2018pib}.  
In Fig.~\ref{fig_FF} we compare the  harmonic number result for $A(K)$ in
(\ref{A2}) (solid curves) to the gluon lattice  results (red-dots). Since the holographic construction does not fix $A(0) $ , we fixed it to 
$A(0)=0.58$ from  the lattice data. The upper green-solid curve is fixed by the Regge mass $m_T(0)=1.386$ GeV, the lower blue-solid curve is fixed by
the Regge mass $m_T(0)=1.330$ GeV. The dashed-purple curve is the dipole approximation 
%It is gratifying that the ensuing holographic result $D(0)=4C(0)=-2.32$ is consistent with the lattice measurement.
%The exact gravitational form factor (\ref{A2}) is well approximated by the dipole form factor

\be
\label{A7}
A(K)\approx \frac {A(0)}{\bigg(1+\frac {K^2}{\tilde m_T^2}\bigg)^2}
\ee
with  $\tilde m_{T}=1.124$ GeV. Throughout and for simplicity, we will use $\tilde m_{T,S}$ for the tensor and scalar parametric masses in the approximate
but accurate dipole fit like (\ref{A7}) to the  holographic result (\ref{A2}). The latter  resums  the unapproximated  tensor and scalar glueball Regge  masses  
$m_{T,S}(n)$ in ({\ref{A3}).
The reported gluon lattice datta are shown in red-dots and well fitted  by a similar dipole with $\tilde m_{T,\,\rm lattice}=1.13$ GeV
which is undistinguishable from the dashed-curve in this mass range~\cite{Shanahan:2018pib}.  $A(0)=0.58<1$
 reflects on the gluon fraction assigned to the nucleon mass, fixed on the lattice but not in holography. We note that a dipole ansatz was originally used 
 in~\cite{Frankfurt_2002} to describe $J/\Psi$ production through two massive gluons close to treshold.

 The approximate dipole form
 (\ref{A7}) from both holography  and the lattice, suggests a tensor radius

\be
\label{A8}
\langle r^2_{GT}\rangle\approx \frac {12}{\tilde m_T^2}\approx (0.62\,{\rm fm})^2
\ee
which is compatible with (\ref{A6}-\ref{A6X}).  
It is comforting that both in our case and the  gluon lattice case, $B(K)$ is consistent with zero. In holography, this is explained by the absence of a tensor
coupling to a bulk Dirac fermion.

(\ref{A2}) and its approximate dipole form (\ref{A7})  resum  the infinite tower of monoples stemming from the full 
$T=2^{++}$ radial Regge trajectory in the dual limit of large $N_c$ and strong coupling. It is rather surprising that this resummed form is comparable
to the  unquenched tensor  form factor  probed by the gluon lattice form factor. This is suggestive of two things:
1/ the quark mixing in the tensor channel is weak; 
2/ the resummed  radial $T=2^{++}$ Regge trajectory  is not very sensitive to the 
lower tensor glueball mass.

\begin{figure}[!htb]
%\minipage{0.6\textwidth}
%  \includegraphics[width=\linewidth]{AK.pdf}
 % \caption{A really Awesome Image}\label{fig:awesome_image1}
%\endminipage\hfill
%\minipage{0.6\textwidth}
% \includegraphics[width=\linewidth]{BK.pdf}
 % \caption{A really Awesome Image}\label{fig:awesome_image2}
%\endminipage\hfill
\minipage{0.8\textwidth}%
  \includegraphics[width=\linewidth]{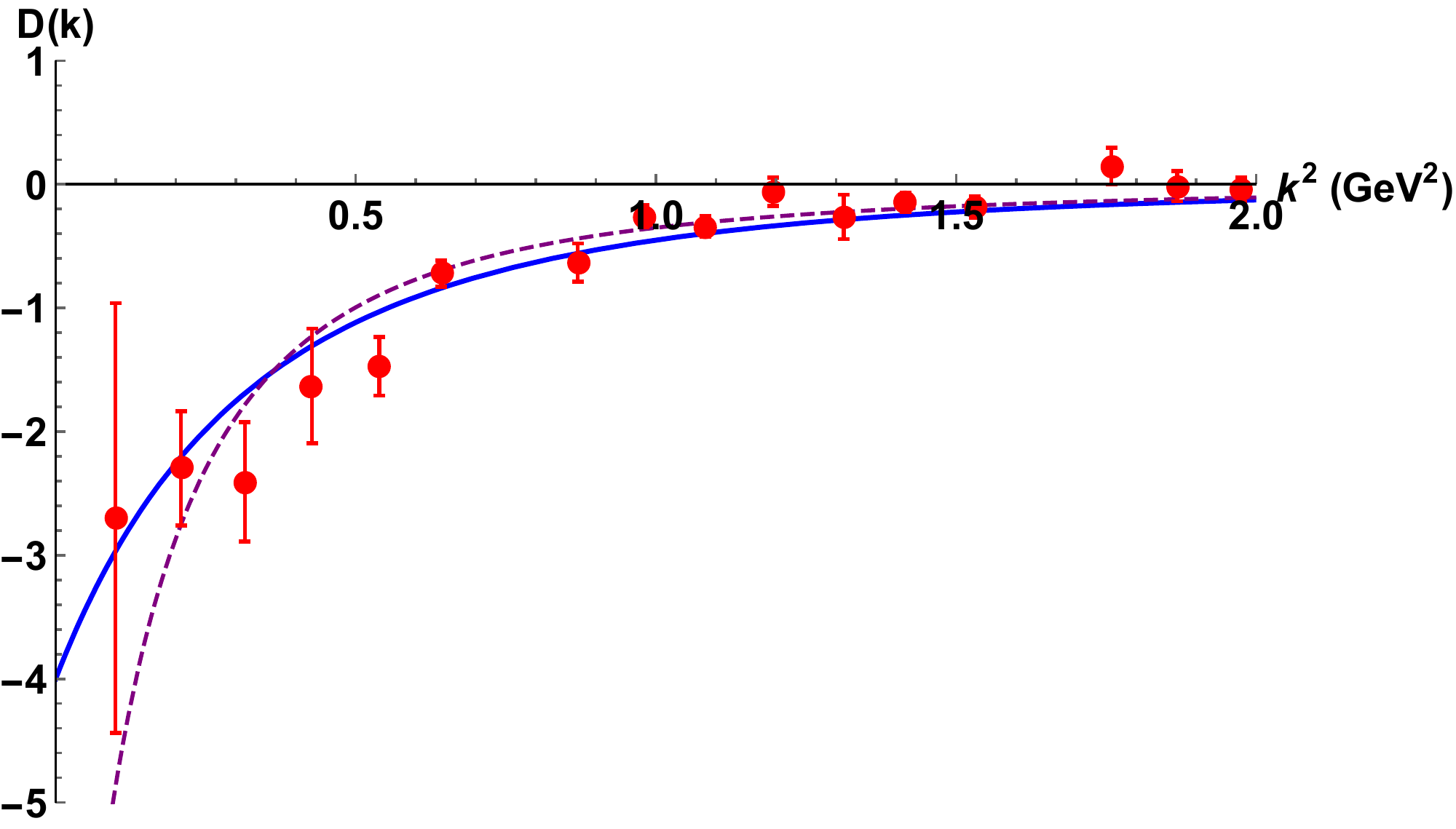}
 % \caption{A really Awesome Image}\label{fig:awesome_image3}
\endminipage
 \caption{Holographic  nucleon gravitational form factors  D(k)=4C(k) (blue-solid curve~(\ref{A7X1})) versus the unquenched   gluon
 lattice results  (red-dots) and lattice fit (purple-dashed curve (\ref{A7X3}))~\cite{Shanahan:2018pib}. See text.}
 % : A(k) top, B(k) middle and C(k) bottom.}
 \label{fig_FFD}
\end{figure}

To compare the D-term with the gluon lattice results, we fit the exact harmonic number results for  both $A(K), A_S(K)$  following from (\ref{EMT2}) ,  with dipoles

\bea
\label{A7X1}
D(K)=4C(K)\approx &&-\frac{4m_N^2}{3K^2}\Bigg[\frac {A(0)}{\bigg(1+\frac {K^2}{\tilde m_T^2}\bigg)^2}- \frac {A(0)}{\bigg(1+\frac {K^2}{\tilde m_S^2}\bigg)^2}\Bigg]\nonumber\\
=&& -\frac{8A(0)}{3}\bigg(\frac{m_N^2}{\tilde m_S^2}-\frac{m_N^2}{\tilde m_T^2}\bigg)
\Bigg[\frac{1+\frac {K^2}{4}\bigg(\frac 1{\tilde m_S^2}+\frac 1{\tilde m_T^2}\bigg)}{\bigg(1+\frac {K^2}{\tilde m_T^2}\bigg)^2\bigg(1+\frac {K^2}{\tilde m_S^2}\bigg)^2}\Bigg]\nonumber\\
=&& D(0)
\Bigg[\frac{1+\frac {K^2}{4}\bigg(\frac 1{\tilde m_S^2}+\frac 1{\tilde m_T^2}\bigg)}{\bigg(1+\frac {K^2}{\tilde m_T^2}\bigg)^2\bigg(1+\frac {K^2}{\tilde m_S^2}\bigg)^2}\Bigg]
%\Bigg[\frac {A(0)}{\bigg(1+\frac {K^2}{m_T^2}\bigg)^2}- \frac {A(0)}{\bigg(1+\frac {K^2}{m_S^2}\bigg)^2}\Bigg]
\eea
with $A(0)=0.58$, $D(0)<0$ and $\tilde m_T\approx 1.124$ GeV.  The scalar mass satisfies

\be
\label{A7X2}
\frac{\tilde m^2_S}{\tilde m_T^2}=\bigg(1-\frac{3D(0)}{8A(0)}\frac{\tilde m_T^2}{\tilde m_N^2}\bigg)^{-1}
\ee
and for a null D-term,  it matches the tensor mass. The D-term falls faster than the A-term and asymptotes $D(K)\approx D(0) (\tilde m_S^2\tilde m_T^2)/K^6$,
which is consistent with the hard QCD counting rules  for the proton D-term~\cite{Tanaka_2018,Hatta:2018ina,Hatta:2021can,Tong:2021ctu}, in sharp 
contrast to the hard QCD scattering rules for the pion D-term~\cite{Shuryak:2020ktq}.
In Fig.~\ref{fig_FFD} we show (\ref{A7X1}) as a solid-blue curve versus the gluon lattice data red-dots,  and the dipole lattice fit 

\be
\label{A7X3}
D_{L}(K)\approx -\frac {10}{\bigg(1+\frac {K^2}{m_D^2}\bigg)^2}
\ee
with $m_{D}=0.48$ GeV  as the dashed-purple curve. The holographic parameters in (\ref{A7X1}) are set to  $D(0)=-4$ and  $(\tilde m_T,\tilde m_S)=(1.124,1.00)$ GeV.
 Again, it is remarkable that the Reggeized holographic result fits rather well the reported lattice data. The latters are unquenched
simulations, and one would have expected  strong 
 scalar-isoscalar quark mixing to the $0^{++}$ glueball states. In particular to the light sigma meson with a  mass of about $0.5$ GeV (although this state is rather broad).   %The holographic result is in this 
% sense the  filtered $0^{++}$ Reggeized gluonic contribution. It is remarkable that the 

The D-term (\ref{A7X1}) allows for the extraction of a new mass radius

\be
\label{AZ1}
\left<r^2_{GD}\right>=-\frac{42}4\bigg(\frac 1{\tilde m_S^2}+\frac 1{\tilde m_T^2}\bigg)\approx -(0.87\,{\rm fm})^2
\ee
which is larger in magnitude than the tensor mass radii (\ref{A6}-\ref{A6X},\ref{A8}), but substantially  smaller in magnitude  than   the lattice  dipole estimate  (\ref{AZ2})

\be
\label{AZ2}
\left<r^2_{GD}\right>_{\rm lattice}=-\frac{12}{m_D^2}\approx -(1.44\,{\rm fm})^2
\ee

We note that since  the $(\tilde m_T,\tilde m_S)=(1.124,1.00)$ GeV   fit to the lattice D-term implies $m_S(n)\neq m_T(n)$ at finite $1/N_c$ in (\ref{A3}), it follows that the 
scalar and tensor radii are  different, with the scalar radius

\be
\label{AZ3}
 \left<r^2_{GS}\right>=\frac{12}{\tilde m_S^2}\approx (0.7\,{\rm fm})^2
\ee
slightly larger than the tensor radii (\ref{A6}-\ref{A6X},\ref{A8}). Recall that  the Reggeized scalar glueballs are sourced by $T^\mu_\mu\approx F^2$
(conformal anomaly) on the boundary,
so the empirical identification of $\tilde m_S$ through the gluonic contribution $D(K)$ in Fig.~\ref{fig_FFD} is justified. The value of $\tilde m_T$ is fixed solely by holography (\ref{A2},\ref{A7}),
and is  consistent with both
the lattice gluonic tensor form factor $A(K)$  in Fig.~\ref{fig_FF}, and the empirically extracted tensor form factor from the  current GlueX  data (see Fig.~\ref{fig_AGlueX} below).

\begin{figure}[!htb]
\minipage{0.8\textwidth}
  \includegraphics[width=\linewidth]{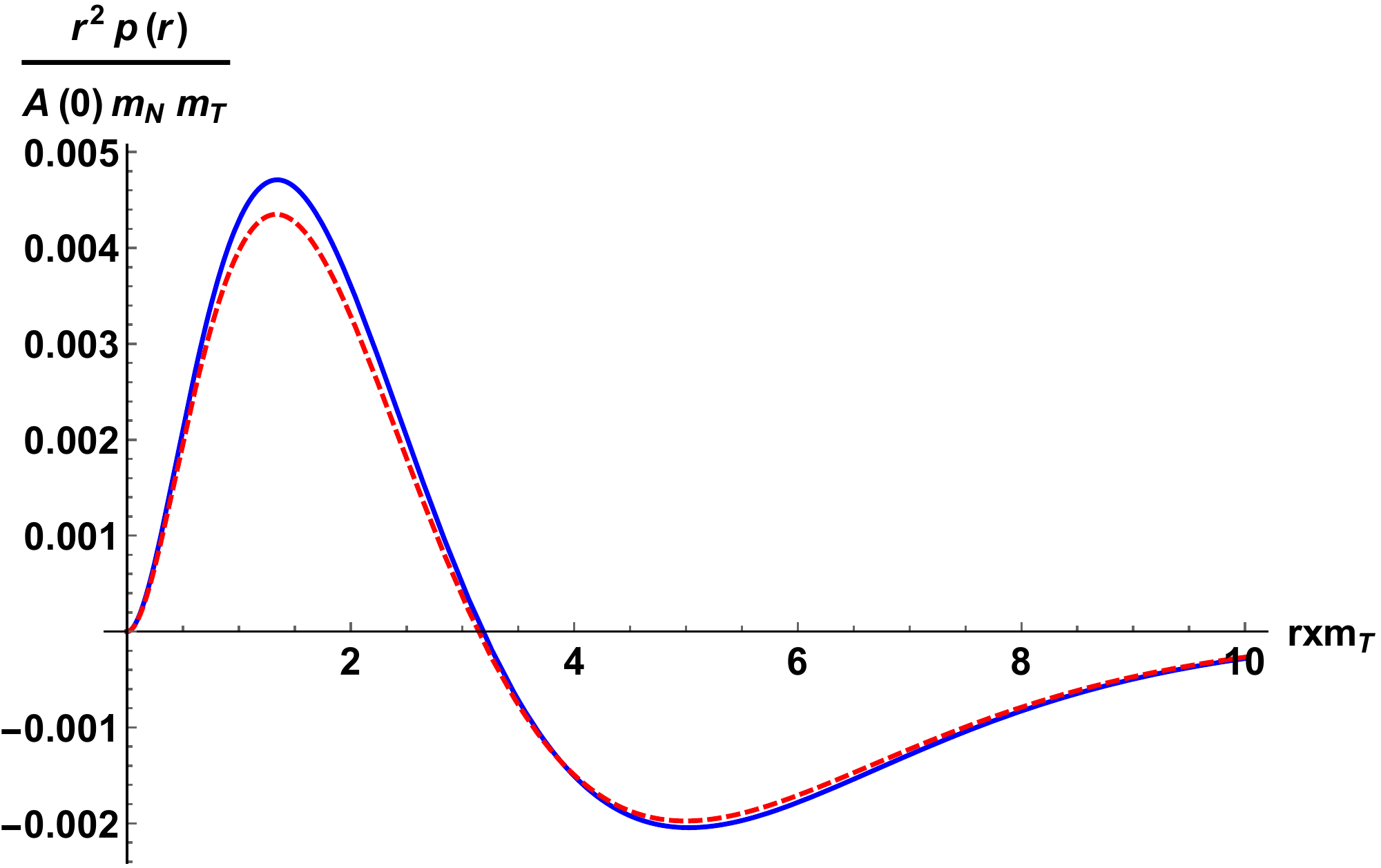}
 % \caption{A really Awesome Image}\label{fig:awesome_image1}
\endminipage\hfill
\minipage{0.8\textwidth}
  \includegraphics[width=\linewidth]{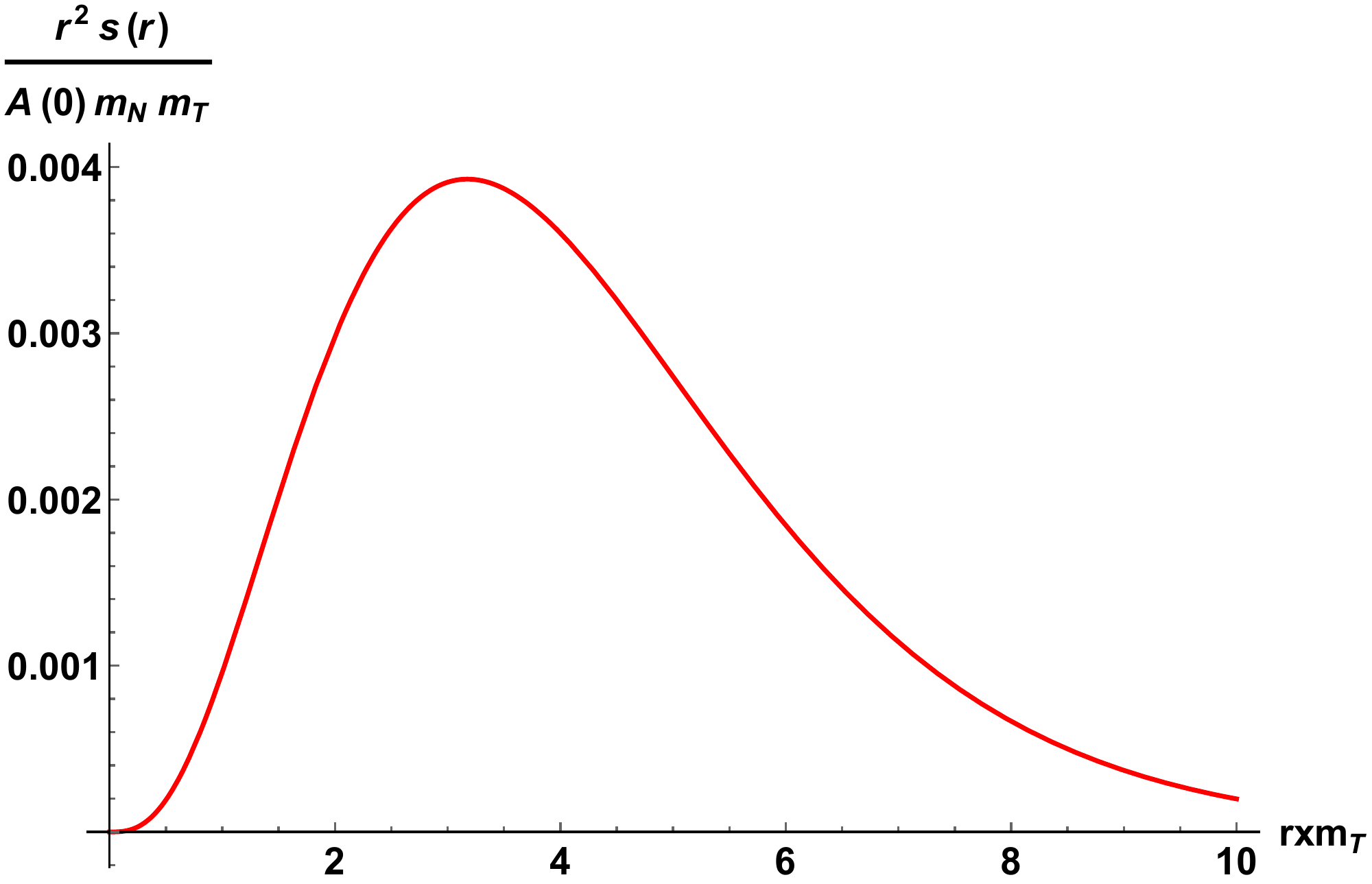}
 % \caption{A really Awesome Image}\label{fig:awesome_image2}
\endminipage\hfill
 \caption{Holographic gravitational pressure and shear inside the proton from (\ref{A9}-\ref{sp}) with $\tilde m_S/\tilde m_T=1/1.124$. See text.}
 % : A(k) top, B(k) middle and C(k) bottom.}
 \label{fig_SP}
\end{figure}

\section{Pressure and shear inside the proton}~\label{PRESSURE}

We follow~\cite{Polyakov:2018zvc} (and references therein) and define the Fourier transform of the D-term

\bea
\label{A9}
\tilde{D}(r)=&&\int \frac{d^3K}{2m_N(2\pi)^3}\,{e^{-i{K}\cdot{r}}}\,D(K)
=\frac{A(0)m_N}{6\pi r}\bigg(e^{-\tilde m_Sr}(2+\tilde m_S r)-e^{-\tilde m_T r}(2+\tilde m_T r)\bigg)\nonumber\\
\eea
where the  approximate but numerically accurate dipole form (\ref{A7}),  instead of the exact harmonic number form (\ref{A2})  for $A(K), A_S(K)$  is used for numerical convenience. 
The   pressure $p(r)$  and shear $s(r)$ distributions in the proton say in the Breit frame, 
can be expressed in terms of $\tilde{D}(r)$ as~\cite{Polyakov:2018zvc} 

\bea
\label{sp}
p(r)=&&\frac{1}{3} \frac{1}{r^2}\frac{d}{dr} \bigg(r^2\frac{d}{dr}
{\tilde{D}(r)}\bigg)\nonumber\\
s(r)= &&-\frac{r}{2} \frac{d}{dr}\bigg( \frac{1}{r} \frac{d}{dr}
{\tilde{D}(r)}\bigg)
\eea
as they capture the anisotropic spatial content of the energy momentum tensor

\be
\label{TIJ}
T^{ij}(\vec r)=\frac 13 \delta^{ij} p(r)+\bigg(\hat{r}^i\hat{r}^j-\frac 13 \delta^{ij}\bigg)s(r)
\ee

In Fig.~\ref{fig_SP}  we show our holographic results  for the radial pressure (top blue-solid curve)
and radial shear (bottom red-solid curve) mass distributions inside the proton from (\ref{sp})
for $\tilde m_S/\tilde m_T=1/1.124$. The scale resolution is fixed at the nucleon mass. The  red-dashed curve in the radial pressure is the estimate

\be
\frac{r^2 p(r)}{A(0) m_N\tilde m_T}\approx \frac {(\tilde m_T r)^2}{200}\bigg(10\,e^{-\tilde m_Tr}-7\,e^{-\tilde m_S r}\bigg)
\ee

The pressure distribution inside the proton is a delicate balance between the Reggeized scalar glueball $S=0^{++}$ attraction,
and the Reggeized tensor glueball $T=2^{++}$ repulsion. The inside of the pressure is dominated by the repulsive tensor, while
the outside of the pressure is dominated by the attractive scalar which ultimatly keeps the proton together. 
The radial shear distribution inside the proton  follows from a much more
subtle difference,  with the scalar and tensor roles reversed.  The Reggeized scalar glueball $S=0^{++}$ contributes positively to the
shear, while the Reggeized tensor glueball $T=2^{++}$ contributes negatively. The difference  is 
in favor of the positive scalar contribution, with a net proton shear positive at all distances.  Note that the scalar contribution to the shear
follows from the traceless constraint on the spatial tensor contribution in (\ref{TIJ}) (quadrupole), which may explain the sign flip.

These results are comparable to the experimentally extracted quark contributions in~\cite{Burkert:2018bqq} for both the pressure
and shear, and also some model calculations in~\cite{Polyakov:2018zvc,Panteleeva:2021iip} (and references therein).
They are also comparable to the  lattice QCD results reported  in~\cite{Shanahan:2018nnv} at the higher scale resolution
$\mu=2$ GeV.

\section{Tensor gravitational form factor from GlueX}~\label{GlueXDATA}

We now show that the
threshold photo-production of heavy vector mesons $V=J/\Psi, \Upsilon$, is solely driven by the tensor gravitational form factor $A(K)$
near threshold, which resums the tensor glueball $T=2^{++}$ radial Regge trajectory. Way above threshold, the same process  is  dominated by
the  Reggeized  form of this form factor following the resummation of the higher  spin-j Regge trajectories leading to the strong Pomeron exchange~\cite{Mamo:2019mka}. 
The A-form factor is measurable  modulo a minimal kinematic assumption which involves the universal coupling of the graviton  in bulk. It  gives access to not only the
proton gluon mass radius, but also the D-term and therefore the gluon shear and pressure distributions inside the proton using holography.

\subsection{Gluon contribution to the proton mass and GlueX}

In QCD the energy-momentum tensor receives contributions from both the quarks and gluons. Its forward matrix element 
in a proton state is fixed by Poincare symmetry

\bea
\label{A10}
\langle P|T_{G+Q}^{\mu\nu}|P\rangle =2(A_{SG}(0)+A_{SQ}(0)=A_S(0)=A(0))P^\mu P^\nu\equiv 2P^\mu P^\nu
\eea
with each of the gluon and quark contributions to the scalar  form factor (\ref{EMT22}) fixed 

\bea
\label{A11}
A_{SG}(0)=&&\frac{\langle P|- {bg^2}F^2/{32\pi^2}|P\rangle}{2m_N^2}\nonumber\\
A_{SQ}(0)=&&\frac{\langle P|m\overline\psi\psi|P\rangle}{2m_N^2}=\frac{\sigma_{\pi N}}{m_N}
\eea
with $b=11 N_c/3-2N_f/3$ (1-loop).
The gluonic  contribution is fixed by the conformal anomaly, and the quark contribution by the pion-nucleon sigma term $\sigma_{\pi N}\approx 50$ MeV.
Note that Poincare symmetry in (\ref{A10}) implies that the tensor glueball exchange and the scalar glueball exchange matches at treshold.
 They both probe the scale anomaly, since $A(0)\approx A_{SG}(0)$  modulo the pion-nucleon sigma term which is small, i.e. $\sigma_{\pi N}/m_N\approx 1/20$.

The differential cross section for photo-production of a heavy meson $\gamma p\rightarrow V p$ 
with $V=J/\Psi, \Upsilon$ can be calculated using the same Witten diagrams as for the form factor, with the result~\cite{Mamo:2019mka}

\bea
\label{A12}
\left(\frac{d\sigma}{dt}\right)
=&&\frac{e^2}{64\pi (s-m_N^2)^2}\times
\bigg[\frac 12 \frac{\kappa^2}{g_5^4}\mathbb V_{hAA}^2 \bigg]\nonumber\\
&&\times\bigg[\frac{A^2(K)}{4m_N^2}\times 
F(s,t=-K^2,M_V,m_N)\times(2K^2+8m_N^2)\bigg]
\eea
Note that asymptotically $F(s,t)\approx s^4$ which implies 
that the corresponding scattering amplitude is ${\cal A}(s,t)\approx s^2$, which is the signature of a $T=2^{++}$ as a graviton
exchange. This growth is tamed by a j-spin Reggeization giving rise to the strong Pomeron exchange at large $\sqrt{s}$. Since the combination

\be
\label{A13}
\bigg[\frac 12 \frac{\kappa^2}{g_5^4}\mathbb V_{hAA}^2 \bigg]\rightarrow \frac 12 \frac{4\pi^2/N_c^2}{(12\pi^2/N_c)^2}\mathbb V_{hAA}^2
\ee
is non-universal, it is unfortunately not possible to extract the value of $A(0)\approx A_{SG}(0)$ from the interpolated differential cross section at threshold.
It is however possible to do more, and extract the full tensor  form factor  as  we have originally suggested and shown in~\cite{Mamo:2019mka}. We now recall  how.

\begin{figure}[!htb]
\includegraphics[height=8cm]{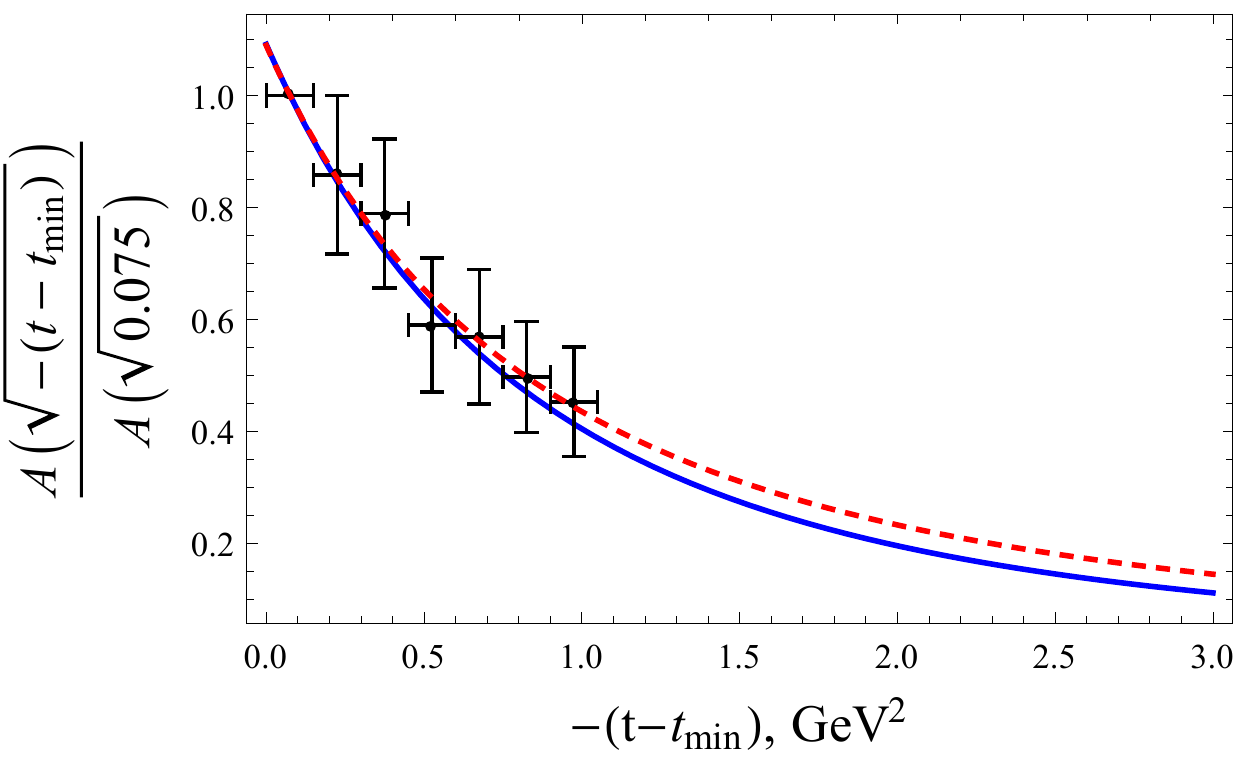}
  \caption{Extraction of the gravitational form factor  $A(\sqrt{-(t-t_{min})})$ (normalized by $A(\sqrt{0.075})$) from the recent GlueX data (crosses)~\cite{Ali:2019lzf}.
  The blue-solid line is the the holographic tensor gravitational form factor  (\ref{A7}) with $m_T=1.124$ GeV and $m_{J/\psi}=3.10~{\rm GeV}$~\cite{Mamo:2019mka}.
  The red-dashed line is the lattice tensor gravitational form factor (\ref{A7}) with $m_{T{\rm lattice}}= 1.13$ GeV~\cite{Shanahan:2018pib}.}
  \label{fig_AGlueX}
\end{figure}

\subsection{Extracting the tensor gravitational form factor from GlueX}

The recently reported GlueX data allows for the extraction of the tensor part of the gravitational form factor
$A(K)$ modulo the kinematical function $F(s,t)$. The latter is fixed by the graviton coupling in bulk
to a spin $1^{--}$ flavor gauge field,  which is  universal.
With this in mind,   (\ref{A12})  suggests to use the empirical ratio of
differential cross sections measured by the GlueX collaborattion~\cite{Ali:2019lzf} to extract $A(K)$. 
In Fig.~\ref{fig_AGlueX} we show this empirical  ratio $A(\Delta t)/A(\Delta t_{\rm min})$ normalized
by  the first or minimal data point, with $\Delta t=(-(t-t_{\rm min}))^{\frac 12}$ and $\Delta t_{\rm min}=\sqrt{0.075}$  
versus $\Delta t^2$ in GeV$^2$~\cite{Mamo:2019mka}

\bea
\label{ATA0}
\frac{A(\Delta t)^2}{A(\Delta t_{\rm min})^2}=
\bigg(\frac{ F(s,t=t_{\rm min},M_V,m_N)(-2t_{\rm min}+8m_N^2)}{ F(s,t=-K^2,M_V,m_N)(2K^2+8m_N^2)}\bigg)
\frac{\bigg(\frac{d\sigma}{dt}\bigg)}{\,\,\,\,\,\,\,\bigg(\frac{d\sigma}{dt}\bigg)_{{\rm min}}}
\eea
The holographic result is shown as the blue-solid line and the lattice result as the red-dashed line.  The
t-dependence in $F(s,t)$ in the range currently probed by GlueX is weak, making the ratio of the differential
cross sections commensurate with the squared  tensor form factor.
The agreement of the extracted GlueX form factor with both calculations suggest that the gluonic part of the shear
and pressure distributions as well as the mass radius is well captured by the holographic 
construction, and now measured. The empirical errors for the ratio have been added in quadrature using the GlueX data. 

The consistency of the GlueX data with the  holographic and lattice estimated curves  using  the dipole  form (\ref{A7}), 
suggests a common tensor mass radius (\ref{A8}) or

\be
\label{RG2}
\langle r^2_{\rm GlueX}\rangle \approx (0.62\,{\rm fm})^2
\ee
A recent  analysis of the same data at threshold extracted  $\langle r^2_{G}\rangle=(0.55\pm 0.03\,{\rm fm})^2$~\cite{Kharzeev:2021qkd} 
which is slightly smaller than the mass radius from the global extraction (\ref{RG2}),  but close to the holographic 
threshold  value (\ref{A6}) from the exact form factor (\ref{A2}) using the simultaneous fit to the nucleon and rho meson
radial Regge trajectories. Finally,  also a  recent analysis of the combined data  for photo-production of
vector mesons, yields $\langle r^2_{G}\rangle \approx (0.64\pm 0.03\,{\rm fm})^2$~\cite{Wang:2021dis}, which  is comparable to (\ref{A8},\ref{RG2}).

\section{Conclusions}~\label{CONCLUSION}

The gluonic content of the three  gravitational form factors is accessible from holographic QCD in the double limit of large $N_c$
and strong coupling.  The A-term is dominated by the tensor $T=2^{++}$ Reggeized radial glueball  trajectories, while the D-term  involves
the difference between the tensor $T=2^{++}$ and $S=0^{++}$ Reggeized radial glueball trajectories, 
which are degenerate in the strict holographic limit (same anomalous dimension).  This degeneracy is lifted in $1/N_c$.
The B-term vanishes in the absence of a tensor coupling to a Dirac fermion in bulk. 

The holographic result for the A-term and D-term compares well with the unquenched lattice results suggesting that there is little quark mixing
in both the $T=2^{++}$ and $S=0^{++}$ channels  when they are fully Reggeized. This is a remarkable observation that can be further tested by
carrying the lattice  calculations in the unquenched limit. The D-term allows for the extraction of both the gluonic pressure and shear inside 
the proton in the holographic limit. We recall that  the holographic scale in the form factors,  follows from the dilaton potential which is 
fixed by the nucleon and rho radial Regge trajectories.

The radial pressure distribution inside the proton follows from a delicate balance between the Reggeized tensor $T=2^{++}$ repulsion
at short distances, and the attraction from the Reggeized scalar $S=0^{++}$ at larger distances. The latter is what ultimatly keeps the
proton together. 
The roles are reversed in the radial distribution of the shear inside the proton, but the trade is much more subtle. 
The net proton   shear stems from  the difference between the scalar sheer due to the $S=0^{++}$  glueball
exchange which is positive, and the tensor  sheer due to the 
$T=2^{++}$ glueball exchange which is negative. The difference  is net positive at all distances

The recently reported GlueX data can be used to extract the full tensor form factor or A-term modulo its threshold normalization.
The tensor mass radius extracted from the GlueX data  $\langle r^2_{\rm GlueX}\rangle  \approx (0.62\,{\rm fm})^2$ using a global fit, 
 is compatible with  $\langle r^2_{GT}\rangle \approx (0.57-0.60\,{\rm fm})^2$ from the holographic threshold result.  
A comparison of the  holographic  D-term to the lattice data, suggests a slightly larger scalar mass radius
$\langle r^2_{GS}\rangle \approx (0.7\,{\rm fm})^2$, among other things. A larger scalar radius was also recently noted in~\cite{Ji:2021}.
 
 Finally, since the holographic  A-term and D-term are related, and  the A-term is accessible from the  GlueX data over a broad range of momenta, 
 we conclude that the GlueX data allows us to glean to the gluonic pressure and shear inside the proton via holography.

\vskip 1cm
{\bf Acknowledgements}

We thank Xiang-dong Ji, Zein-Eddine Meziani and Lubomir Pentchev for  discussion. K.M. is supported by the U.S.~Department of Energy, Office of Science, Office of Nuclear Physics, contract no.~DE-AC02-06CH11357, and an LDRD initiative at Argonne National Laboratory under Project~No.~2020-0020. I.Z. is supported by the Office of Science, U.S. Department of Energy under Contract No. DE-FG-88ER40388.

%\appendix

\bibliography{radius}

\end{document}